\documentclass[lettersize,journal]{IEEEtran}
\usepackage{amsmath,amsfonts,amssymb}
\usepackage{algorithmic}
\usepackage{algorithm}
\usepackage{array}
\usepackage[caption=false,font=footnotesize]{subfig}
\usepackage{textcomp}
\usepackage{stfloats}
\usepackage{url}
\usepackage{verbatim}
\usepackage{graphicx}
\usepackage{cite}
\usepackage{booktabs}
\usepackage{subcaption}
\usepackage{balance}
\usepackage{xcolor}

\usepackage{tikz}
\usetikzlibrary{positioning, fit,patterns, patterns.meta}
\usetikzlibrary {arrows.meta}
\usetikzlibrary{decorations.pathreplacing}
\usepackage{siunitx}

\begin{document}

\title{Physics-Informed Neural Network Models for EMT Simulators}

\author{Ignasi Ventura, Nicolae Darii, Petros Aristidou, Mohammad Kazem Bakhshizadeh, \\ Rahul Nellikkath, Benjamin Vilmann, and Spyros Chatzivasileiadis.
\thanks{\noindent This work was supported by the ERC Starting Grant VeriPhIED, Grant Agreement 949899, and the ERC Proof of Concept PINNSim, Grant Agreement 101248667, both funded by the European Research Council. \\ This work was also supported by the European Union as part of ADOreD project funded by the Horizon Europe MSCA programme (HORIZON-MSCA-2021-DN, Grant agreement 101073554). \par
Ignasi Ventura, Nicolae Darii, Rahul Nellikkath, Benjamin Vilmann, and Spyros Chatzivasileiadis are with the Department of Wind and Energy Systems, Technical University of Denmark, Kgs. Lyngby, Denmark. \par
Petros Aristidou is with the Sustainable Power Systems Lab, Cyprus University of Technology, Limassol, Cyprus. \par
Mohammad Kazem Bakhshizadeh is with Ørsted Wind Power A/s, Fredericia, Denmark.
}
}
\markboth{}%
{Shell \MakeLowercase{\textit{et al.}}: A Sample Article Using IEEEtran.cls for IEEE Journals}

\maketitle

\begin{abstract}
This is the first paper, to the best of our knowledge, to propose a framework that integrates Physics-Informed Neural Network (PINN) models in Electromagnetic Transient (EMT) simulations. Both industry and EMT simulation tools face emerging challenges, as the share of inverter-based resources in power grids rapidly grows. Vendors need to protect the Intellectual Property (IP) of their control design when sharing their component models with customers and system operators. EMT simulation tools need to increase computation speed while maintaining numerical stability. NN surrogates offer numerically stable models with higher degrees of IP protection, as they do not expose the underlying model structure and control parameters. We show that NN surrogates can accurately capture the dynamic behavior of both open-loop and closed-loop controllers, producing identical results in both a validated in-house EMT solver and commercial tools, such as PSCAD. Compared with the artificial delay methods used by commercial solvers, we show that NN surrogates are more numerically robust. Compared with iterative numerical solvers, we show that NN surrogates replacing components with algebraic loops (e.g., closed-loop controllers) increase computation speed by 40\% on average. We demonstrate our methods on PSCAD and the IEEE 39-bus system. Although the speed advantage vanishes when NN surrogates replace only a few components in large scale systems, NN-based models continue to offer enhanced IP protection and numerical stability. Our source code, in-house EMT solver, and the code for seamless integration to PSCAD are made publicly available.
\end{abstract}


\section{Introduction}
%
%
%
%
%

With the rapidly increasing share of Inverter-Based Resources (IBR) in power systems, time-domain simulations based on the Root Mean Square (RMS) approximation are no longer sufficient \cite{milanostability, stabilityupdate}. Power systems with high IBR penetration must consider the electromagnetic phenomena that RMS simulations neglect. As a result, Electromagnetic Transient (EMT) simulations, which reproduce the full waveforms and can duplicate dynamics across all frequencies, are becoming the new standard. At the same time, simulations need to accommodate IBR models that are confidential, as it becomes essential for vendors to protect their Intellectual Property (IP) to maintain competitiveness. With traditional modeling and simulation tools under severe pressure, the power system community seeks (i) computationally efficient and numerically robust methods for time-domain simulations and (ii) methods that can protect the models' IP when shared across the industry.

Modern EMT simulation tools have significantly advanced compared to the first EMT-type simulators~\cite{emtpbookog, jeanEMT}. The programs have undergone substantial improvements in both computing hardware and methodologies~\cite{emtaccelerationreview}, as well as in the core algorithm. 
The simulators now tailor the solution algorithms to the small set of electric components ubiquitous in all power-system models, such as resistors or current sources, and leverage the symmetric nodal structure of electric networks using advanced sparsity, factorization, and decomposition techniques \cite{rtparallel, paraemt, manapaper}.
This significantly accelerated the network solution. When it comes to the solution algorithms for control systems, though, which are abundant in IBR-dominated grids, these targeted improvements are not as obvious. The reason is structural. The solution algorithms need to remain open to accommodate a wide range of control blocks and user-defined models, which constrains the use of tailored numerical techniques that reduce computation time. The primary technique used so far to accelerate their solution is to insert fictitious delays to break feedback loops~\cite{emtpbook, simultaneous2}. This avoids iterative Newton-like methods and is acceptable for well-behaved cases, but produces a non-simultaneous solution that can lead to wrong operating points and numerical instability (see also, for instance, Fig.~\ref{fig:instabilitydelay} of this paper) \cite{araujo}. At the same time, the increased use of encrypted models to protect the vendor IP introduces an additional layer of complexity: the solution algorithm lacks access to the underlying equations, limiting equation-level optimization and slowing down simulations.

Considering the significant progress on Neural Networks (NNs) used for scientific computing over the past few years, NNs emerge as a promising tool able to tackle both of these challenges in EMT simulations. NNs can both (i) ensure enhanced IP protection, as they do not expose, by design, the underlying control system structure or its parameters (i.e. ``black box''), and (ii) offer a numerically stable solution which can also substantially accelerate computations that require numerical iterations, as NNs replace them with a forward matrix multiplication. 

So far, several recent works have sought to leverage the strengths of both NNs and Physics-Informed NNs (PINNs) (i.e. if the underlying equations are included in their training \cite{raissi}) for RMS simulations~\cite{reviewpinns}. In~\cite{chatzivasileiadis2020pinns}, the authors introduced PINNs in the power system literature, proposing to capture all the dynamics of an entire power system. This can provide very fast simulations but is only able to capture a few variables. To scale the speed and modeling advantages of these NNs to large systems, two main approaches have been proposed. The first focuses on NNs for specific components or areas that can modularly integrate into conventional RMS simulation frameworks~\cite{plugandplay,bossart,powertechnew,neuralodes}. Alternatively, \cite{pinnsim} proposed a new simulator concept that models all dynamic components of the system using NNs. While all these works introduce new frameworks for leveraging NNs to accelerate RMS simulations, their applicability to EMT simulations remains an open question. EMT simulations require higher modeling detail and accuracy, and consist of a more complex, targeted solution structure. This makes the entry price for NN-based applications to EMT simulations very expensive.

This paper defines a NN formulation that enables, for the first time to our knowledge, a modular integration of NNs into EMT simulations and commercial tools. Through that, we can overcome numerical stability challenges of existing methods, accelerate EMT simulations, and integrate NNs to existing models in the industry.
The trained NN models, which can also naturally protect the IP of the control design, accurately capture the solution of the chosen control systems, and allow for a new integration method that competes with traditional methods in accuracy, numerical stability, and speed.
In this paper, we demonstrate the seamless integration of NN-based surrogates to an open-source EMT solver and the commercial tool PSCAD simulating the IEEE 39-bus system.

The contributions of this paper are as follows:
\begin{enumerate}
    \item We design Neural Network (NN) surrogate models that accurately capture the solution of control system blocks, accelerate simulations by avoiding numerical iterations, and naturally protect the control design's Intellectual Property (IP).
    \item We mathematically formulate the integration of NN models into existing EMT solvers: we enable NN evaluation as part of standard EMT simulation workflows, and we enhance the EMT solver's numerical stability by avoiding numerical time delays.
    \item We demonstrate how to incorporate these NN models into commercial software, specifically PSCAD, by creating custom-made components whose architecture and evaluation routines are explicitly defined and made available online \cite{github_PINN}. 
\end{enumerate}

For the purpose of this paper, we have created our own EMT solver (thoroughly benchmarked against PSCAD), where we validated our models before implementing them on the PSCAD software. We have made the code of the developed NNs and our EMT solver open-source and publicly available online~\cite{github_PINN}. 

The remainder of this paper is structured as follows. Sec.~\ref{sec2:sims} reviews the EMT simulation program and defines where NNs can bring value into the current framework. Sec.~\ref{sec3:neuralnetsint} explains how NNs and PINNs are formulated and integrated into the existing EMT programs. Numerical results are presented in Sec.~\ref{sec4:numericaltests}, focusing on the conceptual integration, and Sec.~\ref{sec5:pscadimplementation}, focusing on PSCAD simulations with NNs. A discussion and conclusions are presented in Secs.~\ref{sec6:discusssion} and \ref{sec7:conclusions}.

\section{EMT Simulation Workflow and Opportunities} \label{sec2:sims}
In this section, we present the EMT simulation problem and the framework used to develop our in-house EMT solver, and explain how and why NN-based methods can be integrated to accelerate the control-systems solution and improve the privacy of its models.

\subsection{Generic formulation}
Time-domain simulations are defined by two parts: a system model and a time-stepping algorithm. The system model characterizes how the system behaves, typically consisting of a system of nonlinear differential-algebraic equations (DAEs) that represent the physics and control of all components involved \cite{ivp_daes}. The time-stepping algorithm numerically integrates the system model step by step until the simulation reaches the final time. The system model has the following form:
\begin{equation}\label{eq:TDS}
    \mathbf{F}(\Dot{\mathbf{X}}(t), \Dot{\mathbf{Y}}(t), \mathbf{X}(t), \mathbf{Y}(t), \boldsymbol{\eta}, t) = 0,
\end{equation}

where $\mathbf{X}(t)$ represent the electric network states and $\mathbf{Y}(t)$ represent the device models states. These vectors are partitioned as
\[
\mathbf{X}(t)=
\begin{bmatrix}
\mathbf{x}(t)\\
\mathbf{r}(t)
\end{bmatrix},
\qquad
\mathbf{Y}(t)=
\begin{bmatrix}
\mathbf{y}(t)\\
\mathbf{u}(t)
\end{bmatrix},
\]
where \(\mathbf{x}(t)\) and \(\mathbf{y}(t)\) are differential and \(\mathbf{r}(t)\) and \(\mathbf{u}(t)\) are algebraic variables. Array $\boldsymbol{\eta}$ captures the parameters present in the network and the device equations.

To solve this set of equations over time, the time-stepping algorithm iteratively approximates the evolution of the system's differential states from a current value $\mathcal{W}_{t-\Delta t} = \{ \mathbf{X}_{t-\Delta t}, \mathbf{Y}_{t-\Delta t} \}$ to the future value $\mathcal{W}_t = \{ \mathbf{X}_{t}, \mathbf{Y}_{t} \}$ constrained by the system's algebraic equations.

\subsection{EMT Simulation Tool}
The EMT simulation tool aims to represent, in the time domain and in great detail, the dynamics of the modeled power system across a wide frequency range. To do so, the tool captures the instantaneous values of the voltage and current waveforms, being able to capture both electromechanical and electromagnetic phenomena across the system. Eqs.~\eqref{eqs:basicelements} describe the lumped circuit elements, resistor $R$, inductor $L$, and capacitor $C$, across two nodes \textit{k} and \textit{m}. These equations define the voltages $v(t)$ and currents $i(t)$ across the electric network, including both the grid and device-level electric components, which become the state variables $\textbf{X}(t)$ in \eqref{eq:TDS}. The control systems, represented by $\textbf{Y}(t)$ and generally defined with block diagrams, represent the controllers associated with the connected devices.
\begin{subequations} \label{eqs:basicelements}
    \begin{align}
        v_k(t) - v_m(t) &= R i_{km}(t) \\
        v_k(t) - v_m(t) &= L \frac{d}{dt} i_{km}(t) \\
        C \frac{d}{dt} \big( v_k(t) - v_m(t)\big) &= i_{km}(t).
    \end{align}
\end{subequations}

\subsection{EMT Simulation Program} \label{secII-C:EMTP}
There are several approaches to solve an EMT simulation problem. In this work, we consider the partitioned algorithm, widely used in state-of-the-art power system software \cite{emtpbook}. This approach divides the EMT simulation problem into two parts: the electric network and the control systems.

\subsubsection{Electric network}
\begin{figure}[!ht]
    \centering
    \subfloat[Inductance equivalent\label{fig:companion_inductor}]{
        \includegraphics[width=0.45\linewidth]{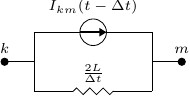}
    }
    \hfill
    \subfloat[Capacitance equivalent\label{fig:companion_capacitor}]{
        \includegraphics[width=0.45\linewidth]{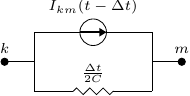}
    }
    \caption{Companion circuits of an inductance and a capacitance.}
    \label{fig:companionmodels}
\end{figure}

The electric network solution is based on the principles outlined by H.W. Dommel in \cite{dommel, emtpbookog}. The differential equations of the circuit elements in \eqref{eqs:basicelements} are transformed into linear algebraic equations by discretizing them with the trapezoidal rule. The resulting equations define the so-called companion circuits and depend on voltage, current, and past values. Fig.~\ref{fig:companionmodels} depicts the companion circuits of an inductance and a capacitance, which use a resistance and a current source to define their behavior. See \cite{emtpbook} for more details on the transformation.

The companion equivalents of all grid elements and the electric circuits of all connected devices are used to build the network equations using Kirchhoff's current law. Assembled together, they yield the network nodal equation:
\begin{equation}\label{eq:nodalequation}
    \textbf{G} \textbf{v}(t) = \textbf{i}(t) - \textbf{I}_{km}(t-\Delta t),
\end{equation}
where $\textbf{v}(t)$ are the bus voltages, $\textbf{i}(t)$ the branch currents, and $\textbf{I}_{km}(t-\Delta t)$ the historical currents at the previous step. $\textbf{G}$ is the nodal conductance matrix. Nonlinear electric elements, such as transformers or surge arresters, can also be included in this formulation \cite{emtpbook}.

We implement and solve \eqref{eq:nodalequation} using linear algebra. However, most commercial solvers nowadays exploit the symmetric and sparse structure of the nodal conductance matrix to develop faster numerical techniques. The most common techniques are the modified nodal analysis \cite{mnapaper}, the modified augmented nodal analysis \cite{manapaper}, and the sparse tableau approach \cite{sparsetableau}.

\subsubsection{Control systems}
The control systems of all devices are solved separately from the electric network. Typically represented by block diagrams, these include converter controls, excitation systems of synchronous machines, protective relays, and many other devices and phenomena that cannot be represented with network elements. The solution algorithm, first proposed in \cite{initacs}, develops all control blocks in the time domain and discretizes them using the trapezoidal rule. The resulting algebraic system then classifies its equations by their nature. Linear equations are combined and solved simultaneously. Nonlinear equations are divided into two groups, depending on whether they present nonlinearities in closed-loop configurations or not. If they do not, they are ordered into a computational sequence and solved analytically. If they do, two main solving approaches are used: (i) introducing a one-time-step delay to break the feedback loops and transform the closed-loop systems into an open-loop ones, or (ii) applying an iterative method to obtain a simultaneous solution.

\begin{figure}[!t]
    \centering
    \subfloat[Closed-loop system.\label{fig_a:tacs_loop_iter}]{
        \includegraphics[width=0.9\linewidth]{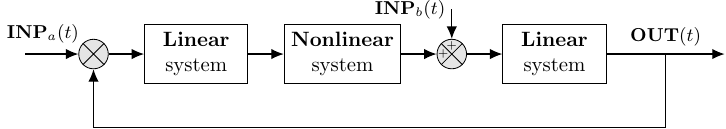}
    }
    \vspace{.6em}
    \subfloat[Delaying the feedback $\mathbf{OUT}(t)$ results in an open-loop system.\label{fig_b:tacs_loop_delay}]{
        \includegraphics[width=0.9\linewidth]{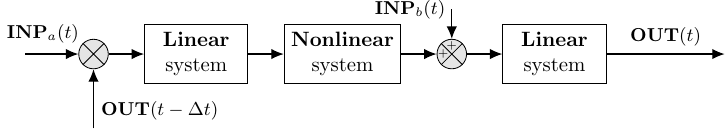}
    }
    \caption{The two main approaches to solve control systems.}
    \label{figs:tacs_loops}
\end{figure}

The most accurate approach is to use an iterative method and obtain a simultaneous solution to the underlying equations, i.e., all variables satisfy the closed-loop interconnection at the same time. A closed-loop system example is shown in Fig.~\ref{fig_a:tacs_loop_iter} with two sets of generic inputs and an output set. Widely used methods include the Newton-Raphson method and its variations. In every iteration, the entire system needs to be repeatedly evaluated, computing the gradients to update the system variables until they all converge. A simultaneous solution, therefore, entails a high computational cost that is often too expensive for simulators to afford.

To avoid iterative methods, solvers typically introduce numerical one-time-step delays into the feedback loops; see Fig.~\ref{fig_b:tacs_loop_delay}. The introduced delays break these loops, allowing the system to be solved analytically in a sequential order and providing a nonsimultaneous solution to nonlinear systems. This approach is significantly faster than the first one and can scale to larger systems. The modeled system is perturbed by a forced, non-physical delay, which alters its dynamic behavior. In many well-behaved cases, these delays do not significantly affect the system's solution. However, especially for fast-varying dynamics, these delays can lead to inaccurate results and even induce numerical instability~\cite{araujo, simultaneous2}.

\subsubsection{Interface network-control solutions}
\begin{figure}[!t]
    \centering
    \includegraphics[width=0.9\linewidth]{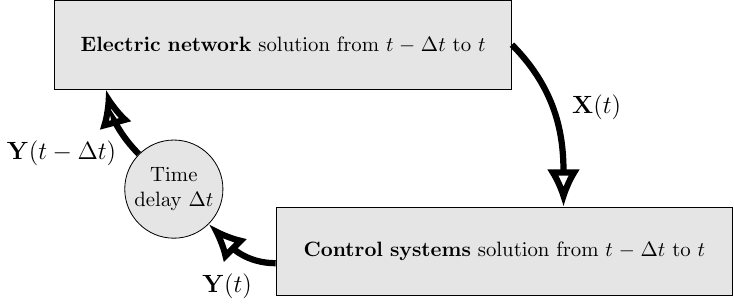}
    \caption{Interface between the network and control systems solution \cite{emtpbookog}.} 
    \label{fig:EMTloop}
\end{figure}

The exchange of information between the electric network and control systems solutions is done sequentially; thus, the network solution always uses the control system's outputs delayed by one time step or vice versa. 
Fig.~\ref{fig:EMTloop} depicts this structure. 
The network is solved from $t-\Delta t$ to $t$ using the control values from the past step, $t-\Delta t$. Then, the control systems are calculated at $t$ with the current electric solution.

\subsection{Capturing Control Blocks with NNs}
We propose a NN formulation compatible with the current control systems solution algorithm, targeting a modular and seamless NN integration, to achieve two key objectives:
\begin{enumerate}
    \item Acceleration of the control systems computation: NNs offer a fast and accurate solution to closed-loop systems, as they (i) require only a forward matrix multiplication instead of numerical iterations, and (ii) are more numerically stable than inserting one-time-step delays.
    \item Preserving the confidentiality of the control system design: NNs do not explicitly expose the underlying control-system structure or its parameters. Once trained, NNs can be shared without revealing the original model equations, parameters, or training data, thereby reducing the need for advanced encryption mechanisms and mitigating risks associated with IP exposure. The information made available is typically limited to the network architecture and a set of fixed weights and biases, from which reconstructing the original model is generally challenging. Furthermore, NN-based models are largely software-independent, facilitating their deployment and exchange across different simulation platforms.
\end{enumerate}

\begin{figure}[!b]
    \centering
    \includegraphics[width=0.99\linewidth]{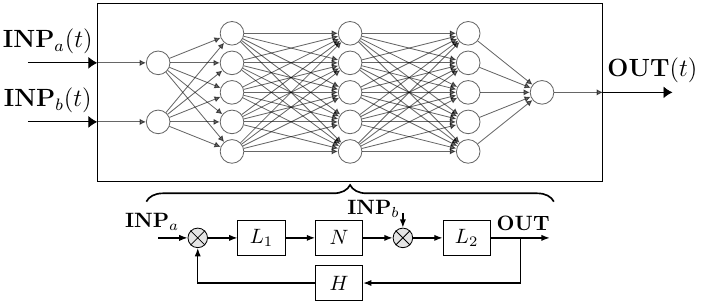}
    \caption{Proposed NN structure. NNs can accurately capture the solution of nonlinear open- and closed-loop control systems.}
    \label{fig:closedloopwrapper}
\end{figure}

Fig.~\ref{fig:closedloopwrapper} illustrates the proposed NN structure. The formulation considers the same two sets of inputs and the set of outputs used in Fig.~\ref{figs:tacs_loops} and aims to learn the mapping from inputs to outputs of the underlying control system.
Sec.~\ref{sec3:neuralnetsint} details how NNs are trained and integrated into the conventional EMT simulation framework.

\section{Neural Networks Integration} \label{sec3:neuralnetsint}
This section defines Neural Networks (NNs) and Physics-Informed Neural Networks (PINNs) and establishes them as an explicit nonlinear modeling method, able to accurately and efficiently capture complex dynamics.
\subsection{General Definition}
We use a fully connected feed-forward NN with $K$ hidden layers and $N_k$ neurons in each layer \cite{Goodfellow}. Each hidden layer $k$ comprises the weight matrices $W^k$ and bias vectors $b^k$. Together, they define the set of learnable parameters of the NN, $\Theta = [{W^k, b^k}]$. The relation between inputs and outputs is defined by 
\begin{equation}\label{eq:hiddenlayer}
    z_{k+1} = \sigma (W^{k+1} z_k + b^{k+1}), \; \forall k = 0, 1, ..., K-1,
\end{equation}

with $\sigma$ being the nonlinear activation function considered in all layers. The combination of learnable parameters and nonlinear activation functions provides them with high expressive capacity. Thereby, NNs can be used to accurately capture complex, nonlinear functions, and even serve as universal function approximators \cite{cybenko}.


\subsection{Training Procedure and Physics-informed Models}
To obtain accurate NN models, we determine the weights and biases $\Theta$ that minimize a training loss function. We use a loss function with two components. Firstly, the classic data-based loss, $\mathcal{L}_u$, minimizes the mean squared error between the target values and the NN predictions. The targets $z$ are provided by a dataset, $\mathcal{D}_u$, which contains accurate simulated results for different time step sizes and initial conditions:
\begin{equation} \label{eq:dataloss}
    \mathcal{L}_u  = \frac{1}{D_u} \sum^{D_u}_{i=1} \left\Vert z^{(i)} - \Hat{z}^{(i)} \right\Vert_2^2.
\end{equation}
Additionally, we can include the underlying differential equations of the system $g$ in training with a physics-based loss $\mathcal{L}_p$. The underlying equations can be incorporated into the training to reduce dataset dependence, improve training accuracy, and enhance the NN's interpretability. The physics-based loss can be applied to the differential equations, provided the activation function $\sigma$ is differentiable. NNs trained with an objective function that includes a physics-based loss term can also be referred to as Physics-Informed Neural Networks (PINNs) \cite{raissi}. The physics-based loss $\mathcal{L}_p$ is formulated as:
\begin{equation}
    \mathcal{L}_p = \frac{1}{D_p} \sum^{D_p}_{j=1} 
    \left\Vert \frac{d}{dt} \hat{z}^{(j)} - g(\hat{z}^{(j)}) \right\Vert_2^2.
\end{equation}

The two loss components are combined using the hyperparameter $\alpha$, which controls the influence of the physics term relative to the data contribution. We solve the optimization problem \eqref{eqs:optimproblem} with a gradient-descent algorithm by adjusting the weights and biases of the selected architecture. The optimization runs until the obtained parameters are accurate enough to represent the underlying system.
\begin{subequations} \label{eqs:optimproblem}
\begin{align}
    \min_{\{W^k, b^k\}_{1\leq k\leq K}} \quad &\mathcal{L}_u(D_u) + \alpha \, \mathcal{L}_p (D_p) \label{loss}\\
    \text{s.t.}\quad & \eqref{eq:hiddenlayer}.
\end{align}
\end{subequations}


\subsection{Proposed NN Formulation}
The proposed NN formulation considers all variables required to determine the simultaneous solution of the desired control system. Because the EMT algorithm advances the system step by step, the NN must approximate the one-step solution and be iteratively evaluated at each step. Using the trapezoidal rule to discretize the system, the variables can be grouped in three categories, as given in \eqref{eq:TDS}. First, the control variables $\mathbf{y}$ and $\mathbf{u}$ are represented at the initial time $t-\Delta t$. Second, the electric variables, that come from the electric network solution $\mathbf{X}$, both at the current time $t$ and at the initial time step $t-\Delta t$. Last, the simulation's step size $\Delta t$. All these variables fully determine the control system solution at the next time step $\textbf{y}(t)$ and $\textbf{u}(t)$ that the NN captures. We can also include initial control values and the time step in the last layer of the NN's architecture to improve its consistency when $\Delta t$ = 0 and to establish the initial condition, as shown below in \eqref{eq:pinnformu}. The control approximations $\hat{\mathbf{y}}_{t}$ and $\hat{\mathbf{u}}_{t}$ are then recurrently used in the simulation and the NN at every next time step.
\begin{equation} \label{eq:pinnformu}
    \hat{\mathbf{y}}_{t}, \hat{\mathbf{u}}_{t}  = \mathbf{y}_{t-\Delta t} + \Delta t \, \cdot \, \mathrm{NN}(\Delta t, \mathbf{y}_{t-\Delta t}, \mathbf{u}_{t-\Delta t}, \mathbf{X}_{t-\Delta t}, \mathbf{X}_t).
\end{equation}

\subsection{Input Domain and Integration}
Our goal is to train NN surrogate models of specific components that can be systematically reused across different scenarios, disturbances, and systems. To enable that, we define an input domain that captures the component's one-step solution across a wide range of its potential operating conditions. For example, the training range for the time step size must be within $\Delta t \in [0, \Delta t_{max}]$. The instantaneous value of voltage waveforms must be within $v_{a,b,c} \in [-1.2 V_{nom}, 1.2 V_{nom}]$. And the tracked angle by a Phase-Locked Loop controller (PLL) must be within $\theta_{pll} \in [0, 2\pi )$. These training ranges are determined for all NN inputs needed to define the underlying blocks, as set in \eqref{eq:pinnformu}. The resulting NN models become agnostic to the simulation they are connected to and can provide an accurate approximation of the one-step component solution for virtually all operating points at which they might be evaluated. 

\begin{figure}[b]
    \centering
    \includegraphics[width=0.99\linewidth]{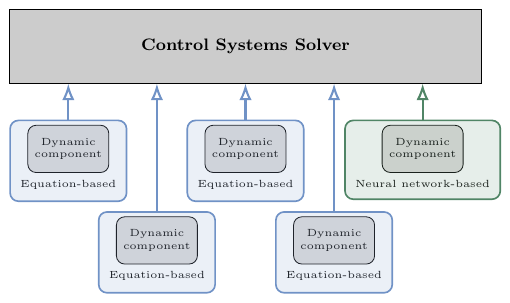}
    \caption{Represented is the NN integration framework for EMT simulations. As many components as required can be seamlessly added in the control systems solver as a NN.}
    \label{fig:int_vision}
\end{figure}

Fig.~\ref{fig:int_vision} illustrates the concept, where in the same system one or more control components can be captured with a NN, enabling a modular and seamless integration.


\subsection{Discussion on Sensitivity and Robustness of NNs}
To provide accurate approximations, NNs require rigorous training with a detailed understanding of the components they capture. One of the main aspects to consider is the input domain: NNs are very accurate at interpolating dynamics within the ranges for which they are trained, but if operating points during a simulation fall outside those ranges, the approximations quickly become inaccurate. Therefore, on the implementation side, it is critical to verify that the queried inputs are within the learned range. A second relevant aspect concerns parameters. The simulation parameters used in training become fixed for deployment, unlike with traditional numerical methods where they can be directly tweaked. Thus, if any of the parameters need to change during simulations, they should be included as NN inputs. Finally, NNs have limits on how much they can learn and the number of inputs and outputs, which are determined by their architecture and training resources. Training balances the NN's capacity with the complexity of the captured dynamics.

\subsection{Training Setup and Implementation}
We implement the NN training and integration using the NumPy and Torch libraries \cite{numpy, torch}. We focus on relatively small architectures with 48-72 neurons and 2-3 layers with a \textit{tanh} activation function. We trained the NNs for $10^6$ epochs with the Adam optimizer and a decaying learning rate. Training was performed on an NVIDIA Tesla A100 GPU \cite{DTU_DCC_resource}. The setup parameters were determined empirically by iteratively tuning the architecture and the training algorithm to achieve the most accurate results on a test dataset. The evaluation of a trained NN becomes the limited set of array operations shown in \eqref{eqs:nnevaluation}. This formulation is written and vectorized in Python and Fortran, and can be quickly computed with CPUs. Once the models are trained, all simulations, with and without NNs, in this paper are run on an i7-1355UCPU@1.70GHz, 16 GB RAM. Although not discussed in this paper, if the EMT solver also supports GPUs, as in \cite{gpusolver1,gpusolver2}, the NN inference becomes faster, leaving room for further acceleration. 
\begin{subequations} \label{eqs:nnevaluation}
    \begin{align}
        x_{a1}   &= \sigma (W^\mathrm{1} x_{} + b^\mathrm{1})\\
        x_{a2}   &= \sigma (W^\mathrm{2} x_{a1} + b^\mathrm{2})\\
        x_{a3}   &= \sigma (W^\mathrm{3} x_{a2} + b^\mathrm{3})\\
        y_{}     &= W^4 x_{a3} + b^\mathrm{4}
    \end{align}
\end{subequations}

\section{Numerical Test Results} \label{sec4:numericaltests}
We investigate the integration of NNs in EMT simulators in two different environments. First, we developed our own EMT solver, which we specifically developed in Python for this work, to enable better insight into the implementation and validation of our proposed framework. We followed the formulation detailed in Sec. \ref{sec2:sims}, and thoroughly validated it against PSCAD to ensure it reproduces the same numerical solutions across a wide range of simulations  (see Appendix for validation results). This in-house EMT solver, open-source and available online in \cite{github_PINN}, provides a complete overview of the simulation and numerical workflow in which we integrate the NNs. Second, we integrated the same NN models directly into the commercial software package PSCAD V5, as components of the IEEE 39-bus system, to demonstrate how the proposed NN surrogate models can be integrated into commercial EMT simulators and existing EMT simulation models.

\subsection{Study Case: Grid-Following Converter} \label{secIV-A:StudyCase}
\begin{figure*}[b]
    \centering
    \includegraphics[width=0.975\linewidth]{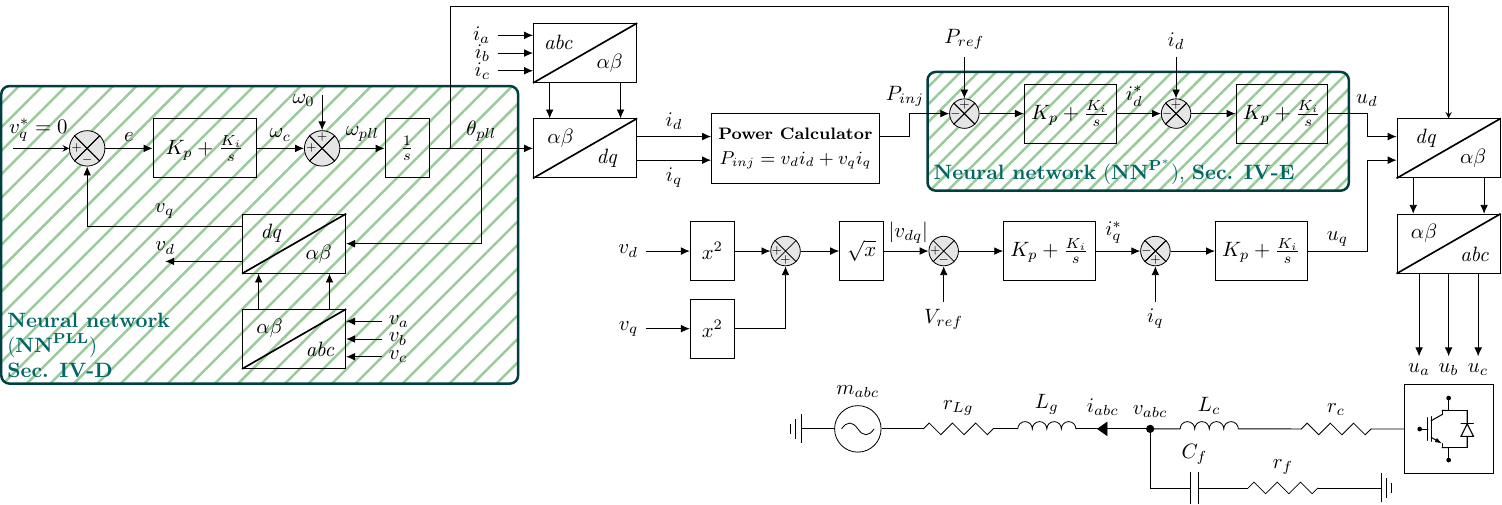}
    \caption{Block diagram representation of the WT control systems. We integrate a NN that captures a PLL, Sec.~\ref{secIV-B:PLLshow}, and the outer and inner active power control loop, Sec.~\ref{secIV-C:Ploopshow}.}
    \label{fig:control_sys_diagram}
\end{figure*}
We consider an aggregated type-4 Wind Turbine (WT) model connected to an external grid through an LCL filter \cite{kazem}. We model the grid-side converter as an average controlled voltage source, assuming a constant DC voltage and neglecting the rotor-side converter, since we focus on the interactions between the grid and the WT under different grid faults. Fig.~\ref{fig:control_sys_diagram} depicts the control architecture of the considered WT model. A Synchronous Reference-Frame PLL controller (SRF-PLL) is implemented to synchronize with the external grid. We apply vector control in the \textit{dq} frame, using a cascaded PI configuration that, in the outer loop, controls the active power injected into the external grid and the voltage magnitude at the Point of Common Coupling (PCC). We consider an ideal modulation block. The system configuration and parameters are based on the CIGRE benchmark model C4.49~\cite{Lukash}. Table~\ref{tab:sysparams} presents the parameters considered for the WT system.
\begin{table}[!h]
\centering
\caption{System and control parameters for aggregated WT}\label{tab:sysparams} 
\begin{tabular}{ccc}
\toprule
\textbf{Symbol} & \textbf{Description} & \textbf{Value}  \\
\midrule
  $S_b$  & Rated power & 100 MW   \\
  $V_g$  & Nominal grid voltage & 100 kV \\
  $f_0$ & Rated frequency & 50 Hz \\
  $T_s$ & Simulation time step size &[1-100] $\unit{\micro\second}$ \\
  $K_{pll}$ & SRF PLL gains (kp, ki) & 25, 300 \\
  $K_{pc}$ & Power controller gains (kp, ki) & 0.5, 30 \\
  $K_{cc}$ & Current controller gains (kp, ki) & 0.1, 20 \\
  $r_c, L_c$ & Converter-side inductor (pu) & 0.005, 0.1 \\
  $r_f, C_f$ & Filter capacitor (pu) & 0.0757, 0.00184 \\
  $r_{Lg}, L_g$ & Grid-side inductor (pu) & 0.005, 0.1 \\
  \bottomrule
\end{tabular}
\end{table}

\subsection{Study Setup}
The results presented throughout this paper rely on only two NNs. One model that captures the PLL solution, $\mathbf{NN^{PLL}}$, and another model that captures the outer and inner control loops for active power, $\mathbf{NN^{P^*}}$. Fig.~\ref{fig:control_sys_diagram} shows which block diagrams are captured by these two NNs. We selected these two control blocks as they are widely implemented examples of closed-loop (PLL) and open-loop (active power control) systems in power systems control. 

We first run the defined study case in the EMT solver we developed, choosing whether to use a NN surrogate or a standard equation-based model for the PLL and active power control loops. Except for the representation of these two control systems, the rest of the model, as well as the system and solver configuration, are identical. This setup enables a very targeted study to compare the performance of the EMT solver when NNs are used instead of equation-based models.

To study their performance, we run multiple simulations for transient stability studies. The simulations consider four types of events: large active-power and voltage-setpoint step changes, grid voltage sags, and grid voltage phase-angle jumps. Sec.~\ref{secIV-B:PLLshow} and Sec.~\ref{sec5:pscadimplementation} review the performance of the $\mathbf{NN^{PLL}}$, while Sec.~\ref{secIV-C:Ploopshow} focuses on the $\mathbf{NN^{P^*}}$.

\subsection{One-Time-Step Delay Approach} \label{sec:onetimestepdelay} 
As explained in Sec.~\ref{secII-C:EMTP}, introducing a one-time-step delay to break nonlinear control feedback loops is a common approach in EMT programs, such as PSCAD, to accelerate the solution of control systems. Although this offers a very fast solution approximation, it can produce inaccurate results and induce spurious numerical instability. 

\begin{figure}[!t]
    \centering
    \includegraphics[width=0.999\linewidth]{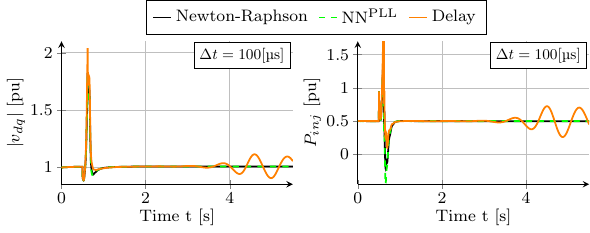}
    \caption{Solving with one-time-step delays can lead to numerical instability, while Newton-Raphson and NN solutions remain stable and accurate.}
    \label{fig:instabilitydelay}
\end{figure}

To better understand how the one-step delay leads to numerical instability, Fig.~\ref{fig:instabilitydelay} presents the active power injection and the voltage at the PCC of the presented study case, following a large-grid-voltage-sag disturbance.
The Newton-Raphson method and the one-time-step delay are applied to capture the PLL system shown in Fig.~\ref{fig:control_sys_diagram}.
Using the same $\Delta t=100 \, \unit{\micro\second}$, Fig. \ref{fig:instabilitydelay} shows how using the Newton-Raphson for the closed-loop PLL control captures the true trajectory for the system, while the delay fails. 

\subsection{Closed-Loop Control Systems} \label{secIV-B:PLLshow}
\begin{figure}[!b]
    \centering
    \includegraphics[width=0.999\linewidth]{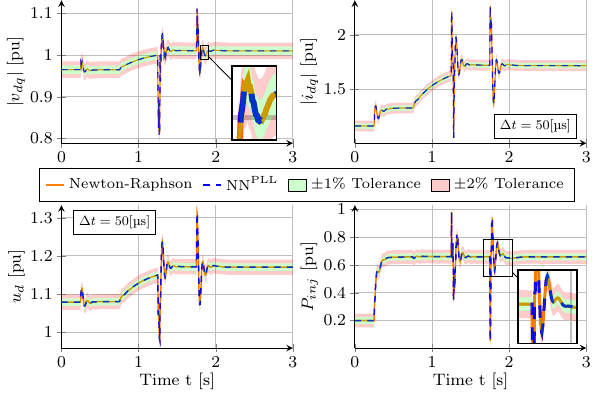}
    \caption{EMT simulation results for four different events with the PLL blocks solved by a Newton-Raphson and the $\mathbf{NN^{PLL}}$.}
    \label{fig:pll_pinn}
\end{figure}

One of the goals of this paper is to introduce NN surrogates that maintain the speed but avoid the numerical fragility of the fictitious one-time-step delay for closed-loop controllers we observed in the previous section. As Fig.~\ref{fig:instabilitydelay} shows, $\mathbf{NN^{PLL}}$ achieves the same accuracy as the Newton-Raphson method and is 60\% faster than the Newton-Raphson method by eliminating the need for iterations.
Diving deeper into the performance of $\mathbf{NN^{PLL}}$ versus the equations-based model using the Newton-Raphson method, Fig.~\ref{fig:pll_pinn} depicts simulation results for a 3-second simulation with four consecutive events, two initial setpoint step changes, with the active power $P^*$ from 0.2 to 0.66 pu and the voltage $V_{ref}$ from 0.97 to 1.01 pu, followed by a grid voltage sag from 0.965 pu to 0.8 pu for 25 $m s$ and a $20^\circ$ phase angle jump. The simulation time step is $\Delta t=50 \, \unit{\micro\second}$. Except for the PLL control loop modeling, the two simulations are identical. Fig.~\ref{fig:pll_pinn} shows how the $\mathbf{NN^{PLL}}$ tracks the angle as accurately as the Newton-Raphson method, with \emph{all} errors bounded within 0.3\% tolerance and resulting in a 2x faster simulation. It is interesting to note that the 100\% acceleration gain we achieve also shows how significantly a single feedback control loop can slow down an EMT simulation.

\begin{figure}[!t]
    \centering
    \includegraphics[width=0.999\linewidth]{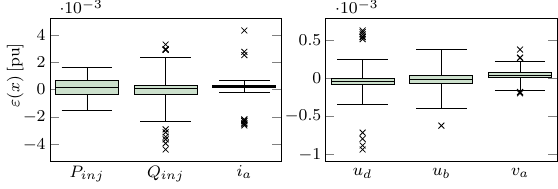}
    \caption{Error boxplots for a two-minute simulation comprising 100 events. The errors are calculated as the difference between using the Newton–Raphson method and the $\mathbf{NN^{PLL}}$ to solve the PLL; otherwise the solvers are identical.}
    \label{fig:boxplots_pll}
\end{figure}

Next, we run a 2-minute simulation with 100 different events of the four studied types, in random order, duration, and magnitude, and demonstrate how our NN integration framework provides a method that remains accurate and numerically stable while significantly accelerating the simulation. Fig.~\ref{fig:boxplots_pll} shows the error boxplots over the 2-minute simulation, presenting the most relevant electric and control variables. We calculate the errors as the differences at each time step between the simulation that solves the PLL using the Newton-Raphson method on the equation-based model and the simulation that uses $\mathbf{NN^{PLL}}$. For the shown 2-minute simulation, implementing $\mathbf{NN^{PLL}}$ results in a 40\% faster simulation on average while accurately capturing the simulation transients. The interested reader can reproduce all results through our code available on \cite{github_PINN}.

\subsection{Open-Loop Control Systems} \label{secIV-C:Ploopshow}
\begin{figure}[!t]
    \centering
    \includegraphics[width=0.999\linewidth]{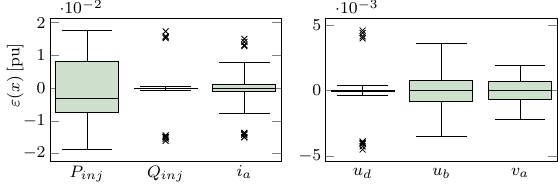}
    \caption{Error boxplots for a two-minute simulation comprising 100 events. The errors are calculated as the difference between using a sequential and $\mathbf{NN^{P^*}}$ to solve the $P^*$ control loops.}
    \label{fig:boxplots_pstar}
\end{figure}

%
From an acceleration or numerical stability standpoint, NN surrogates offer no advantage for open-loop controllers. Existing EMT simulation tools solve these systems fast and accurately, provided they have access to the corresponding mathematical equations. However, if users wish to protect their IP, NNs can help preserve it. Assuming a vendor wishes to protect the IP of an open-loop controller, here we use $\mathbf{NN^{P^*}}$ to blackbox the system's active power controller, and report how this affects the simulation speed and accuracy.
We run the same 2-minute simulations with 100 different events as in the previous section and again focus on the simulation errors between the equation-based solution and the $\mathbf{NN^{P^*}}$ approximations. Fig.~\ref{fig:boxplots_pstar} shows that replacing the active power control with a NN surrogate results to slightly larger (but still acceptable) errors throughout the simulation compared with the replacement of the PLL in the previous section. The reason for that is probably that the active power controller sits much closer to the PCC, and considering that our performance metrics are primarily the voltage, current, and power at the PCC, i.e. $u_a, u_b, u_c, i_a, i_b, i_c,$ etc., similar errors in the NN output can have a slightly larger impact on the final outputs. Still, all errors remain within a 2\% tolerance, and $\mathbf{NN^{P^*}}$ demonstrates an accurate and stable performance. When it comes to solution speed, however, $\mathbf{NN^{P^*}}$ increases the simulation time, on average, by 1.5x compared to the sequential solution. The reason is that the sequential solution consists of only 6 equations, whereas the NN requires a number of matrix multiplications involving its weights and biases. Here, it becomes evident that the NN structure plays a role. Smaller NN architectures can increase the solution speed but might suffer from lower accuracy.


\section{Integration to Commercial EMT Simulator} \label{sec5:pscadimplementation}
This section implements the $\mathbf{NN^{PLL}}$ model into the PSCAD WT model, and demonstrates it on the IEEE 39-bus test system in PSCAD. We validate the system with and without the $\mathbf{NN^{PLL}}$ in both the time- and frequency-domain.

\subsection{PSCAD and Fortran Implementation}
PSCAD runs an EMT solution engine called EMTDC, developed in Fortran. To include NNs in the EMTDC engine, we require two processes. First, a reading function that fetches all NN parameters from a structured storage file. Second, EMTDC shall evaluate the NN model at each step and output the solution to the other control systems and the electric network; through that, the NN becomes another native component in PSCAD. To achieve this, we define a custom component in PSCAD that declares the NN architecture in a Fortran script. This Fortran script loads the NN parameters (e.g., weights and biases) from the data file, and evaluates the associated NN during the simulation. Internal variables are then used to transfer information from one step to the next. This process can be scaled to any number of NNs, enabling the modeling of any type and any number of control systems with NN models. 

\subsection{Wind Turbine Model: Time-Domain Validation}
\begin{figure}[!t]
    \centering
    \includegraphics[width=0.999\linewidth]{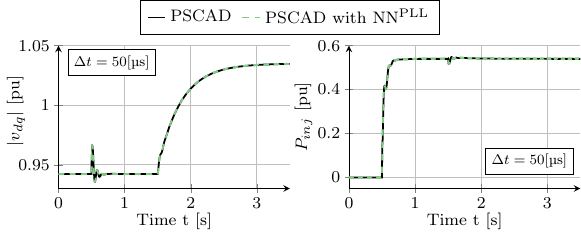}
    \caption{PSCAD trajectories with and without $\mathbf{NN^{PLL}}$ for a large $P^*$ and $V$ setpoint steps.}
    \label{fig:pscad_trajectories}
\end{figure}

Fig.~\ref{fig:pscad_trajectories} presents the trajectories for voltage magnitude and injected active power at PCC comparing the use of the standard PSCAD PLL model with the $\mathbf{NN^{PLL}}$. Similar to the results we showed with our in-house EMT solver, $\mathbf{NN^{PLL}}$ yields highly accurate results in PSCAD, closely matching the default conventional model.
In terms of simulation speed, PSCAD models remain faster than $\mathbf{NN^{PLL}}$ since PSCAD uses the one-time-step delay approach. 
Using the delay approach, PSCAD breaks the PLL closed loop and solves the system as an open loop. As we have seen in Section~\ref{secIV-C:Ploopshow}, EMT simulators can efficiently solve open-loop systems. In contrast, the NN evaluation requires array operations, which increase the number of computations per step and make the simulations 20\%-80\% slower, depending on the NN size. Still, NNs maintain the advantage of IP protection for proprietary control design, which is important for the industry, while offering a method that improves numerical stability. 

\subsection{Wind Turbine Model: Frequency-Domain Validation}
We also validate the PSCAD model with the $\mathbf{NN^{PLL}}$ in the frequency domain through a frequency scanner, a well-established validation tool for EMT models. In contrast to time-domain analysis, a frequency scan systematically probes the system across a broad frequency range and can reveal mismatches that may never appear in a single transient event. If a surrogate model reproduces the resonance frequencies and magnitudes of the equation-based model, this is strong evidence that it captures the underlying dynamics correctly. 

\begin{figure}[!b]
    \centering
    \includegraphics[width=0.999\linewidth]{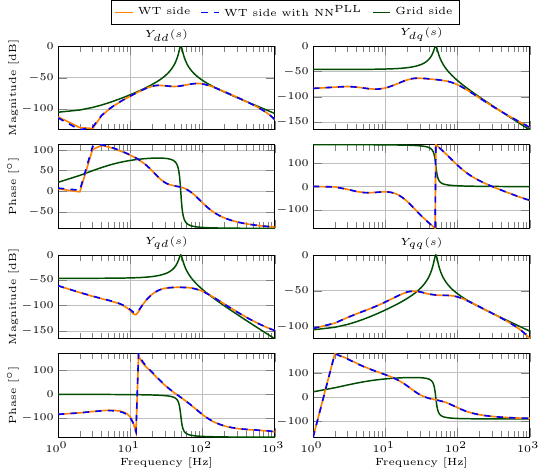}
    \caption{Frequency response of the PSCAD model with and without $\mathbf{NN^{PLL}}$.}
    \label{fig:bodeplots}
\end{figure}

We run a frequency scanner at the operating point of $P_{inj}^0$=50.28 MW and $V_{pcc}^0$=100 kV \cite{freqscanner}. A total of 800 frequency points are sampled between 1 Hz and 1 kHz, enough to cover all the control loops and LCL filter bandwidth. Fig.~\ref{fig:bodeplots} presents the Bode magnitude and phase responses for both configurations.
Both responses on the WT side closely overlap across the frequency range, demonstrating that $\mathbf{NN^{PLL}}$ accurately captures PLL dynamics across its operating range and can be integrated into commercial EMT simulators. The frequency response on the grid side is identical for both configurations because the grid-side representation is the same.

\subsection{IEEE 39-bus System}
\begin{figure}[!t]
    \centering
    \includegraphics[width=0.999\linewidth]{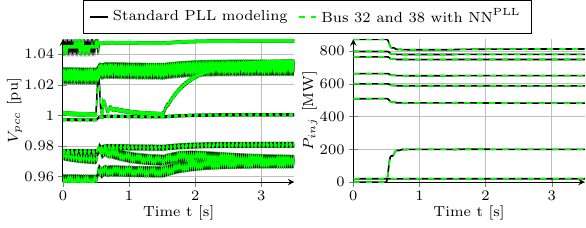}
    \caption{PSCAD trajectories with and without $\mathbf{NN^{PLL}}$ for a large $P^*$ and $V$ setpoint steps at the WT at bus 32.}
    \label{fig:ieee39bus}
\end{figure}

To demonstrate the modularity and scalability of the proposed methodology, we use $\mathbf{NNs^{PLL}}$ to simulate the IEEE 39-bus benchmark system~\cite{39bussys} in PSCAD. We replace two voltage sources at buses 32 and 38 with the presented WT model, resulting in a system with 8 voltage sources and 2 WTs. 
Fig.~\ref{fig:ieee39bus} shows the results after introducing a large active power setpoint step change in the WT at bus 32. 
The configuration with the $\mathbf{NNs^{PLL}}$ yields almost identical results to the standard PLL modeling, demonstrating that NNs can be used to model control systems and be seamlessly integrated into PSCAD. In this larger case, the NNs evaluation cost difference compared with the conventional PSCAD one-time-step delay approach is negligible, as the two PLLs represent a small part of the total control systems solution cost. As a result, if we only plan to replace a limited number of closed-loop controllers with their NN surrogates in large-scale system studies, the main benefit of $\mathbf{NNs^{PLL}}$ is the IP protection of the potentially proprietary control design.

\section{Discussion} \label{sec6:discusssion}
The proposed methodology opens the stage to explore how Scientific Machine-Learning (ML) methods can address some of the challenges faced by both the industry, with the need to protect proprietary control designs, and the EMT simulators themselves with the high computational cost and potential numerical fragility.
We deliver a new formulation that synergizes power system simulations and ML by focusing on the NN strengths and the simulations' bottlenecks. Below, we discuss several topics to consider when implementing NNs as a surrogate model.

\subsubsection{Scaling vs Accuracy}
NN representational capacity is limited by their architecture size and the number of inputs and outputs considered. 
While we can achieve high accuracy with 5-10 variables, increasing the input-output dimensions makes the training challenging and reduces accuracy. Advances in the ML field are expected to dramatically increase the scalability of these models in the near future. Until then, common scaling approaches include using different NN models for different ranges, avoiding internal variables in the input-output sets, or modeling a system with several connected NNs.

\subsubsection{Upfront training cost}
NN training is computationally intensive; the $\mathbf{NN^{PLL}}$ used in this paper required 12 hours. This training time should be considered, however, as a single-time cost, similar to the time necessary to design and fine-tune an equations-based model. Once trained, NNs can be stored in a model library and reused as many times as desired, similar to conventional models. 
Our vision is that commercial simulation tools can soon include in their model library NN surrogate models next to the conventional ones.

\subsubsection{Data-driven against equation-based modeling}
NNs rely entirely on data and are therefore somewhat agnostic to the absolute convergence of the physical systems they model. When the shown simulations with NNs reach a steady state, we observe the outputs oscillating around the equilibrium point. NNs recurrently approximate the next solution based on the evaluated inputs, and therefore, cannot reach a zero-gradient equilibrium point. As long as these oscillations are within a very small range around the true equilibrium point, we consider the results to be correct.

\subsubsection{Modeling from real data} The NN models developed in this paper were driven by physics-informed training and by simulations of existing physics-based models. 
However, the framework we propose enables commercial EMT simulators to seamlessly integrate ML models that are trained directly on real data and represent processes that equation-based models fail to capture (e.g., components with missing/false parameters, demand-side flexible resources, distribution network equivalents, etc.). Developing and examining the performance of such models is a promising direction for future work.

\section{Conclusions} \label{sec7:conclusions}
This is the first paper, to the best of our knowledge, to propose a framework for integrating NN surrogate models into EMT simulators. Our approach addresses challenges faced by both industry and EMT simulation tools. First, considering that NN models offer higher degrees of IP protection for proprietary control designs, equipment vendors can now use the formulation proposed in this paper and share NN-based models that seamlessly integrate with EMT simulators; such models accurately capture the behavior of controllers without the risk of exposing the underlying structure and parameters. Second, this paper showed how PINNs offer numerically robust models that accelerate conventional simulation tools by up to 100\% and avoid the numerical fragility of the one-time-step delay approach used by PSCAD and other commercial simulators to break closed-loop systems. 

We demonstrate the integration of NN surrogates into both a validated in-house EMT solver and the commercial simulator PSCAD, testing it on simulations up to the IEEE 39-bus system. We examine the performance of NN models for both open- and closed-loop controllers using a type-4 wind turbine model. We find that in both simulation tools, the NN surrogates (i) maintain high accuracy, producing almost identical trajectories to the equation-based models, and (ii) avoid numerical instabilities that can appear because of the one-time-step delay in fast-evolving cases. In terms of computation speed, we find that NN models result in a 40\% faster solution on average when replacing closed-loop controllers. In case of replacing open-loop controllers NNs increase the simulation time. In all cases, if users intend to only replace a few components with NN-based models in a large-scale simulation, we do not observe a significant change in the computation speed. 

To enable the integration of NN-based models in existing EMT simulators, the proposed NN formulation, along with the code to integrate NN-based models in PSCAD, and the validated in-house EMT solver are publicly available at \cite{github_PINN}.

\section*{Appendix} \label{sec8:appendix}
\begin{figure}[!h]
    \centering
    \includegraphics[width=0.99\linewidth]{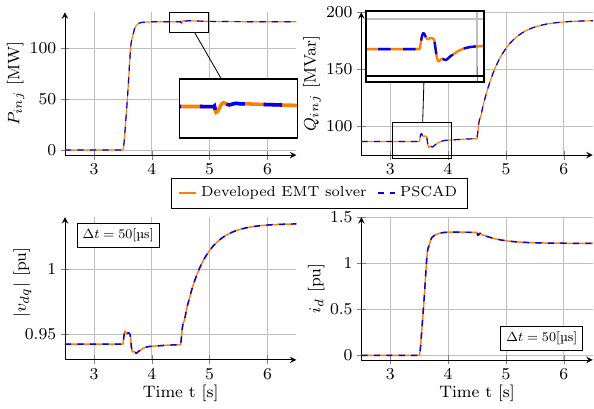}
    \caption{Power, voltage, and current traces of the study case simulated with the developed EMT solver and PSCAD.}
    \label{fig:validation}
\end{figure}

The developed EMT solver used for the numerical results in Sec.~\ref{sec4:numericaltests} is verified by implementing the same study cases in PSCAD. Fig.~\ref{fig:validation} shows the trajectories obtained with the developed solver and PSCAD for a large double setpoint-change trajectory. We apply a $1$ s ramp to increase the active power injected into the grid from 0 pu to 1 pu at $t = 2$ s. Three seconds later, at $t = 5$ s, we introduce a setpoint change of injected reactive power from 0 pu to 0.2 pu. We verify that the dynamics and operating points coincide.

\balance 

\bibliographystyle{ieeetr}
\bibliography{Sections/references}

\end{document}